# OH terminated two-dimensional transition metal carbides and nitrides (MXenes) as ultralow work function materials

Mohammad Khazaei[1*], Masao Arai[1], Taizo Sasaki[1], Ahmad Ranjbar[2], Yunye Liang[3], and Seiji Yunoki[2,4,5]

[1]Computational Materials Science Unit, National Institute for Materials Science (NIMS), 1-1 Namiki, Tsukuba 305-0044, Ibaraki, Japan
[2]Computational Materials Science Research Team, RIKEN Advanced Institute for Computational Science (AICS), Kobe, Hyogo 650-0047, Japan
[3]New Industry Creation Hatchery Center, Tohoku University, Sendai 980-8579, Japan
[4]Computational Condensed Matter Physics Laboratory, RIKEN, Wako, Saitama 351-0198, Japan
[5]Computational Quantum Matter Research Team, RIKEN Center for Emergent Matter Science (CEMS), Wako, Saitama 351-0198, Japan

*Corresponding author. E-mail address: khazaei.mohammad@nims.go.jp

**Abstract:** MXenes are a set of two-dimensional transition metal carbides and nitrides that offer many potential applications in energy storage and electronic devices. As an important parameter to design new electronic devices, we investigate the work functions of bare MXenes and their functionalized ones with F, OH, and O chemical groups using first-principles calculations. From our calculations, it turns out that the OH terminated MXenes attain ultralow work functions between 1.6 and 2.8 eV. Moreover, depending on the type of the transition metal, the F or O functionalization affects increasing or decreasing the work functions. We show that the changes in the work functions upon functionalizations are linearly correlated with the changes in the surface dipole moments. It is shown that the work functions of the F or O terminated MXenes are controlled by two factors: the induced dipole moments by the charge transfers between F/O and the substrate, and the changes in the total surface dipole moments caused by surface relaxation upon the functionalization. However, in the cases of the OH terminated MXenes, in addition to these two factors, the intrinsic dipole moments of the OH groups play an important role in determining the total dipole moments and consequently justify their ultralow work functions.

# 1. Introduction

The work function is one of the basic physical characteristics of a material that is related to the bulk properties (through the Fermi energy) and is influenced by the surface properties such as the surface index, the surface relaxation, and the surface adsorbates [1-11]. Moreover, the work function is one of the most important technological parameters in developing new field emitter cathodes or new Schottky barrier junctions for the applications in the light emitting diodes and the field effect transistors [12,13]. Materials with low work functions are demanded for many electronic applications so as to enhance the power efficiency. A typical example of such materials is cesium, for which the work function is 2.1 eV [1]. It is also known that the cesium plays a crucial role in lowering the work function when it is deposited on various surfaces [14-16]. This is along the general ideas that the deposition of the alkali metals or other elements is capable of tuning the surface dipole moments, which in turn directly affect the work functions [17-22]. However, the usage of the cesium is always accompanied with some major concerns because of its toxicity. Recently, new techniques have been developed successfully to grow self-assembled monolayers of organic molecules with low work functions, which are eco-friendly [23] although they might be unstable thermally.

Various metallic substrates of transition metal carbides and nitrides provide the high stabilities, the high melting points [24], and the relatively low work functions [25-28] that have increasingly attracted attentions as suitable field emitters [29,30]. Their thin films also have already been synthesized experimentally [30]. Remarkably, experimentalists succeeded recently to synthesize the two-dimension monolayers and multilayers of a particular family of transition metal carbides and nitrides with chemical composition of $M_{n+1}C_n$ and $M_{n+1}N_n$ (M is an early transition metal), which were named MXenes [31,32]. The MXenes are derived from exfoliation of the MAX phases [33-37] --- $M_{n+1}AX_n$, where A is an element from groups 13-16 in the periodic table and X is carbon or nitrogen --- using the hydrofluoric acid. During the acid treatment, the A element is removed from the MAX phase and the surfaces of the obtained two-dimensional $M_{n+1}X_n$ are instantaneously terminated by F, OH and/or O groups. The MXenes have already found applications as energy storage materials in Li ion batteries and high volumetric capacitors [38-46], as transparent conductive electrodes [47], filler in polymeric composites [48], purifier [49], dual-responsive surfaces [50], and a suitable substrate for dyes [51]. Theoretically, there are many proposed applications for the monolayers, multilayers, and heterostructures of the MXenes in the electronic [52-64], magnetic [65,66], optoelectronic [67], thermoelectrics [52,68], sensors [69], Schottky barrier junction devices [70-72].

Considering properties of the MXenes, it is noticed that the members of this family can be good candidates as materials with the low work functions by the following reasons: first, the MXenes offer the tunability of the work function by choosing the proper transition metal as well as the X element, usually carbon or nitrogen. In addition, such compositional tunability provides a possibility to adjust other properties such as thermal, mechanical, chemical stabilities, and toxicity. Second, during the synthesis of the MXenes, their surfaces are functionalized naturally, affecting the electronic structure, especially, giving the Fermi level shift and the electrostatic potential change near the surfaces. They are caused by the electron redistribution in energy and in real space, respectively. Both changes have significant effects on the work function. Currently, the MXenes are functionalized by mixture of O, F, and/or OH. From a viewpoint of the development of low work function materials, the OH termination is interesting because the OH group has an intrinsic dipole moment.

In this paper, using first-principles calculations based on the density functional theory (DFT), we propose the functionalized MXenes as materials with the tunable low work functions. Here, we have studied the work functions of a large number of the two-dimensional MXenes with various thicknesses and transition metals such as $M_2C$ and $M_{10}C_9$ (M= Sc, Ti, Zr, Hf, V, Nb, and Ta) as well as $M'_2N$ and $M'_{10}N_9$ (M'= Ti, Zr, and Hf) functionalized with F, OH, and O. As examples of structures of the functionalized MXenes, the model structures of $M_2C(OH)_2$ and $M_{10}C_9(OH)_2$ are shown in Fig. 1. It is predicted from the calculations that the OH terminated carbide or nitride MXenes attain the ultralow work functions between 1.6 and 2.8 eV comparable with the cesium-deposited surfaces. In addition, depending on the M element, the F or O termination affects the work function to increase or decrease as compared with their corresponding bare structures without functionalization.

## 2. Calculation and analysis method:

We used DFT with the Perdew–Burke–Ernzerhof (PBE) version of the generalized gradient approximation (GGA) for the exchange-correlation functional [73] to optimize atomic structures and calculate electronic structures. The basis sets were generated using the projector augmented wave (PAW) method with a cutoff energy of 520 eV. The Methfessel–Paxton smearing scheme was applied with a smearing width of 0.1 eV [74]. The two-dimensional system was modeled by using the supercell approximation with a vacuum space along the direction perpendicular to the surface, which is taken to be the z-axis. The atomic positions and the two-dimensional cell parameters were fully optimized using the conjugate gradient method so that the magnitude of the force acting on each atom became less than 0.001 eVÅ$^{-1}$ in the optimized structures and the total energies were converged within $10^{-7}$ eV/cell. For the structural optimization, the Brillouin zone was sampled using a set of 12×12×1 $k$ points [75]. All calculations were performed using the VASP code [76].

As described above, the exposed surfaces of the MXenes are functionalized with F, OH, and/or O during exfoliation process. We have previously shown that the MXenes with the full surface functionalizations are thermodynamically more favorable than the partial functionalizations [52], where the full surface functionalization requires two chemical groups per cell. The surfaces of the MXenes have two types of the hollow sites: hollows with and without a carbon/nitrogen sites beneath them [52]. Thereby, we tested three initial configurations to find the optimum positions for the attached chemical groups: Structure 1) both chemical groups are attached to hollows without carbon/nitrogen; Structure 2) both chemical groups are attached to hollows with a carbon/nitrogen atom; Structure 3) two chemical groups are attached to two different types of hollows. The optimization results of the above three initial configurations are listed in the supplemental material [77]. We used the lowest energy configurations for the work function analyses. Moreover, for the cases of the OH-terminated systems, we performed additional test calculations by tilting the H atoms toward the nearest transition metals and carbon/nitrogen atoms. It was found in the considered MXenes the OH groups locate always perpendicular to the surfaces. We also calculated the phonon dispersions of some of the OH-terminated MXenes by using Phonopy package [78]. As shown below, the phonon modes are all positive, indicating the stability of such systems.

The work function was derived from the energy difference between the Fermi energy and the vacuum level (electrostatic potential away from the surface). We used an extremely large vacuum space of 80 Å to prevent the interactions between monolayers and their images along the z-axis. The convergence error of the calculated work function is smaller than 1.0 meV (see the supplemental material [77]). The structures are placed in the center of the cell. We validate our computational approach by performing sets of work function calculations on pure transition metals as well as binary transition metal carbides and nitrides [77]. The results were compared with the available theoretical and experimental data [79-91].

By using the Poisson equation, the electrostatic potential away from the surface can be related to the surface dipole moment, and thus, the work function [92]. Leung *et al.* derived an expression relating the change in the work function ($\Delta\Phi$) and the change in the surface dipole moment ($\Delta P$): $\Delta\Phi = -e/\varepsilon_o \Delta P = -180.95\Delta P$, where $\Delta\Phi = \Phi-\Phi_o$, $\Delta P = p-p_o$, and $e$ (C) and $\varepsilon_o$ (CV$^{-1}$ Å$^{-1}$) are the charge of electron and vacuum permittivity, respectively [92]. The $\Phi_o$ ($\Phi$) and $p_o$ (p) are the work function and the surface dipole moment before (after) functionalization, respectively. In order to obtain the surface dipole moments, we used the average charge density ($\rho(z)$) along the *z*-axis, which was obtained from the total charge density $\rho(x,y,z)$ by $\rho(z) = (1/A) \iint dx\, dy\, \rho(x,y,z)$ where *A* is the cell surface area. It is noted that the total charge density includes the contributions of both electrons and ions. The ion charge densities are defined as point charges using the delta function distributions at the positions of the ions. The surface dipole moment can be determined by $p(z) = \int_{z_o}^{z} z'\rho(z')dz'$, where $z_o$ (0.0 Å) is the center of the MXene.

Upon deposition of any element, three main factors control the surface dipole moment: i) the redistribution of electron charge between the surface and the adsorbates, ii) the surface relaxation caused by the adsorbates, and iii) the polarity of the adsobates. Following the theoretical method developed by Leung *et al.* [92], the electron charge transfer ($\Delta\rho(z)$) between the adsorbates and the substrate is defined as $\Delta\rho(z)=\rho(z)-[\rho_s(z)+\rho_a(z)]$, where $\rho(z)$, $\rho_s(z)$, and $\rho_a(z)$ are the total charge densities of the functionalized system, the substrate, and the adsorbate, respectively. Computationally, $\rho_s(z)$ is generated from a total energy calculation of the substrate by removing the overlayer of adsorbates from the functionalized system without the structural optimization. Similarly, $\rho_a(z)$ is obtained from a total energy calculation of the adsorbate layer by removing the substrate from the functionalized system. Thereby, the change in the total dipole moment ($\Delta P$) by functionalization can be expressed as $\Delta P = \Delta p + p_a + p_s - p_o$, where $\Delta p$, $p_s$, and $p_a$ are the induced dipole moments of the $\Delta \rho(z)$, $\rho_s(z)$ and $\rho_a(z)$, respectively. As a consequence, the change in the work function becomes $\Delta \Phi = -(e/\varepsilon_o)(\Delta p + p_a + p_s - p_o)$. Therefore, $\Delta \Phi$ is affected by the dipole moment of the transferred charges ($\Delta p$) between the substrate and the adsorbates, the dipole moment of the adsorbates ($p_a$), and the change in the dipole moments due to substrate relaxation ($p_s - p_o$).

The work function of a functionalized two-dimensional system can be affected by another factor, the shift of the Fermi level. In contrast to the three-dimensional systems, the Fermi energy can shift by the electron transfer caused by the surface functionalization. Therefore, the surface functionalization of thin materials can change both the Fermi energy and the surface dipole moments, but in the bulk three-dimensional materials, the surface functionalization modifies mainly the surface dipole moments.

## 3. Results and discussions:

We have systematically studied the work functions of the MXenes with different M elements: $M_2C$ and $M_{10}C_9$ (M= Sc, Ti, Zr, Hf, V, Nb, Ta) as well as $M'_2N$ and $M'_{10}N_9$ (M'= Ti, Zr, and Hf) functionalized with F, OH, and O. The results of the calculated work functions are shown in Fig. 2. Figure 2 does not include the results for $Sc_2CO_2$, $Ti_2NF_2$, and $Ti_{10}N_9F_2$, in which the functional groups are adsorbed on different types of the hollow sites. These MXenes obtain intrinsic dipole moments due to the asymmetric attachments of the functional groups. Here, we focus only on the MXenes without the intrinsic dipole moments to understand the systematical change of the work functions.

Figure 2 clearly shows that in each MXene family, their thin and thick monolayers manifest similar behaviors upon the same functionalization. The work functions of the bare MXenes and the functionalized ones by F, OH and O are found to be distributed in the range of 3.3-4.8, 3.1-5.8, 1.6-2.8, and 3.3-6.7 eV, respectively. Specially, we found that that $M_2C(OH)_2$, $M_{10}C_9(OH)_2$, $M'_2N(OH)_2$ and $M'_{10}N_9(OH)_2$ exhibit ultralow work functions as compared with their corresponding bare and the functionalized ones with F or O. Among these systems, $Sc_2C(OH)_2$ and $Sc_{10}C_9(OH)_2$ exhibit the lowest work functions of ~1.6 eV. However, depending on the M element, the work function of the functionalized MXenes with F or O can be higher or lower than that of their corresponding bare structures. For instance, the F functionalization on $Sc_{10}C_9$ increases the work function up to 4.99 eV from the bare value 3.35 eV, but on $Hf_{10}C_9$ decreases the work function down to 3.19 eV from the bare value 4.76 eV. The O functionalization increases the work functions for most of the considered MXenes, except for $Hf_{10}N_9$ and $Sc_{10}C_9$. Among the considered MXenes, $V_2CO_2$ and $V_{10}C_9O_2$ possess the highest work functions of ~6.7 eV. The work function reduction upon the F or O functionalization is apparently in contradiction with the common accepted rule that high electronegative elements such as F or O increase the work functions. It is worth mentioning that such unexpected behavior has also been reported in other materials. For example, oxygen adsorption on the tungsten (100) shows the work function reduction [92].

It is noteworthy that MAX phases with n ≥ 3 exist: $Ti_4SiC_3$ [93], $Ti_4AlN_3$ [94], $Ti_4GeC_3$ [95], $V_4AlC_3$ [96], $Nb_4AlC_3$ [97], $Ta_4AlC_3$ [98], $Ta_4GaC_3$ [99], $Ti_5SiC_4$ [93], $(Ti_{0.5}Nb_{0.5})_5AlC_4$ [100], $Ta_6AlC_5$ [101], $Ti_7SnC_6$ [102]. These experimental observations indicate that formations of MAX phases ($M_{n+1}AX_n$) with n ≥ 3 are possible though the production of their single phases may need further optimization of the experimental conditions. Thereby, it is fundamentally interesting to study the thickness dependence of the work functions of the derived $M_{n+1}X_n$ MXenes in detail. Hence, next we investigated the thickness dependence of the work function functionalized by F, OH, and O. As typical examples, we calculated the work functions of $Ti_{n+1}C_n$, $Ti_{n+1}C_nF_2$, $Ti_{n+1}C_n(OH)_2$, and $Ti_{n+1}C_nO_2$ (n = 1-9) and the results are shown in Fig. 3. In these Ti-based MXenes, the thicknesses of the bare ones extend from 2.31 ($Ti_2C$) to 22.15 Å ($Ti_{10}C_9$). As seen in Fig. 3, the thin and thick MXenes act similarly upon the same functionalization: the F or O functionalization increases the work functions, while the OH functionalization decreases it significantly. In general, the dependency of work functions of the MXenes on their thicknesses is weak.

The calculations also indicate that the thickness affects the work functions of the thin MXenes. This is due to the quantum size effects [7-9,103], which result in the shift of the Fermi energy, and consequently the change of the work function. In very thin MXenes, few energy bands cross the Fermi energy. Therefore, any elemental deposition may have significant effects on the electronic structure, i.e., the number of bands and band shapes, and thus the position of the Fermi energy varies (see the supplemental material [77]). The largest change in the work functions of the functionalized $Ti_{n+1}C_n$ occurs between $Ti_2CO_2$ and $Ti_3C_2O_2$. This change is induced because of the specific properties of these systems: $Ti_2CO_2$ is a semiconductor [52] while $Ti_3C_2O_2$ is metallic (see supplemental material [77]). From thickness dependence analyses of the work functions of functionalized $Ti_{n+1}C_n$, it is recognized that the change of the Fermi energy, which in principle affects the value of the work functions, cannot explain the ultralow work functions found in the OH terminated $Ti_{n+1}C_n$.

Since the modifications of the work function of the MXenes upon different functionalizations cannot be justified by the change of the Fermi energy alone, it is necessary to examine the other contributions to the work function, i.e., the electrostatic potential. As described above the electrostatic potential away from the surface can be estimated from the surface dipole moment [92]. In the thick MXenes ($M_{10}C_9$ and $M'_{10}N_9$), the functionalization does not change the Fermi energy significantly. Hence, this allows us to focus on the electrostatic term only (surface dipole moments) to explain the changes in the work functions after functionalization. Figure 4 summarizes the results for the changes in the work functions ($\Delta\Phi$) of $M_{10}C_9$ and $M'_{10}N_9$ and their corresponding surface dipole moments ($\Delta P$) upon different functionalization relative to their bare systems. In Fig. 4, the positive (negative) values of $\Delta\Phi$ correspond to the increase (decrease) of the work function. It is found that $\Delta\Phi$ correlates with $-\Delta P$ almost linearly with the slope value of 170.012 VÅ. The analysis by Leung et al [92] has shown that $-\Delta\Phi/\Delta P = 180.95$ VÅ for the bulk surfaces, and no deviation were observed. In contrast, our results apparently deviate from this relationship. In their analysis, the shift of the Fermi level upon the functionalization was not taken into account because such shift does not occur in the bulk. On the other hand, the functionalization of monolayers can have a non-negligible effect on the Fermi level. Thus, the deviation is due to the finite shift of the Fermi level.

In order to understand how the chemical groups affect the surface dipole moments, we considered in details all terms of $\Delta p$, $p_a$, $p_s$, and $p_o$ contributing to the surface dipole moments ($\Delta P = \Delta p + p_a + p_s - p_o$) in details. As typical examples, we dissolve the total dipole moments into contribution of $\Delta p$, $p_a$, $p_s$, and $p_o$ for systems, $Hf_{10}C_9F_2$, $Sc_{10}C_9F_2$, and $Sc_{10}C_9(OH)_2$. Note that the F functionalization decreases (increases) the work function of $Hf_{10}C_9$ ($Sc_{10}C_9$), while the OH functionalization decreases work function of $Sc_{10}C_9$. The results of $\Delta p$, $p_a$, $p_s$, and $p_o$ for these systems are shown in Figs. 5-7.

Figures 5-7 also include the redistributions of the transferred electron charges between the substrate and the functional groups ($\Delta \rho$). A common feature is that the charge redistributions are mainly localized in the regions nearby the surfaces and the adsorbates. It is also observed that in all MXenes studied here, the natures of F and O elements are the same: due to their higher electronegativity than transition metals and/or hydrogen atoms of OH, these elements always receive electrons from the nearby elements. Thereby, as shown in Figs. 5-7(a), several regions with positive (nearby transition metals and/or hydrogen indicated by ① and ③) or negative charges (nearby the F and O indicated by ②) are formed. The positive or negative charges in regions ①, ② and ③ contribute largely to the $\Delta p$ [92]. As shown in Figs. 5(b) and 6(b), the induced dipole moments $\Delta p$ of the redistributed charges $\Delta \rho$ for $Hf_{10}C_9F_2$ and $Sc_{10}C_9F_2$ becomes positive for $Hf_{10}C_9F_2$ at the distance of 40 Å from the surface while it is negative for $Sc_{10}C_9F_2$ at the same position. This clarifies why the F deposition causes the decrease of the work function in $Hf_{10}C_9F_2$, but the increase in $Sc_{10}C_9F_2$. In fact, the charge transfer has already been attributed to the increase or decrease of the work functions in many materials. For example, the charge transfer explains the effects of different adsorbates on the work functions of graphene or boron-nitride sheets [104-106], and the reduction of the work function for the tungsten (100) surface upon the oxygen deposition [92]. However, the low work function behaviors of the OH terminated MXenes cannot be explained by the effect of $\Delta p$ only. In addition to $\Delta p$, it is necessary to take into account the contributions of the induced dipole moments from surface relaxations ($p_s$-$p_o$) and the adsorbates ($p_a$) in the total dipole moments ($\Delta P$).

The term $p_s$-$p_o$ indicates the modifications in the surface dipole moments due to the changes in the atomic layer distances upon the chemical functionalization, i.e., the structural relaxation effect. In Figs. 5-7 (c), the sharp changes in the $p_s$ or $p_o$ occur at the position of the atomic layers. It is noted in Figs. 5-7 that in order to avoid the complications of the figures, we have shown only the position of the atomic layers in the functionalized MXenes but not in the bare MXenes. As seen in Fig. 5(c) the F functionalization on $Hf_{10}C_9$ does not change the atomic layer distances, and consequently $p_s$-$p_o$ is zero at the distance of 40 Å far away from the surface. However, as seen from Figs. 6(c) and 7(c), upon the F or OH functionalization on $Sc_{10}C_9$, the layer distances near the surfaces change outwardly as compared with the bare surface. Hence, the term $p_s$-$p_o$ contributes to the total dipole moments. The reason why the F or OH adsorbates affect the layer distances of $Sc_{10}C_9$, but not in $Hf_{10}C_9$, might be explained by the layer elastic constants ($C_{33}$) as our $C_{33}$ calculations indicate that $Sc_{10}C_9$ (219.9 GPa) is softer than the $Hf_{10}C_9$ (457.7 GPa) [77].

Figures 5-7(d) show the dipole moment contributions of the adsorbate layers ($p_a$). As expected and also shown in Figs. 5(d) and 6(d), the layers of F or O groups do not have any net dipole moment away from the surface. However, OH is a polar chemical group. Hence, the OH layer has an intrinsic dipole moment as shown in Fig. 7(d). We found that the contribution of the OH dipole moment to the total dipole moments is large and its sign is opposite to the contributions of the other terms ($\Delta p$ and $p_s$-$p_o$). This explains why the OH terminated MXenes exhibit the exceptionally low work functions, in spite of the fact that OH has an electronegative nature similar to F and O groups. It should be noted that the importance of the effect of the intrinsic dipole moments of chemical groups has been pointed out on the increase or decrease of the work functions of the Au (111) surfaces as well. For example, it has been shown that the deposition of alkylthiolates (-$SCH_3$) increases the work function of Au (111), whereas deposition of fluorinated alkylthiolates (-$SCF_3$) decreases it [107].

In order to show the OH terminated MXenes are locally stable structures, we have performed a set of phonon calculations on $Sc_2C(OH)_2$, $Ti_2C(OH)_2$, $Zr_2C(OH)_2$, $Zr_2N(OH)_2$, $Hf_2C(OH)_2$, $Hf_2N(OH)_2$, and $Ta_2C(OH)_2$. The phonon modes of these systems are found to be all real and positive. The phonon dispersions of $Sc_2C(OH)_2$ (the MXene with lowest work function) and $Ti_2C(OH)_2$ (one of the well studied MXenes experimentally [31,32,38,47,108,109]) are shown in Fig. 8. The others are summarized in the supplemental material [77]. Interestingly, in Fig. 8 the phonon branches above 100 THz correspond to the starching modes of OH bonds [109]. The weak dispersion indicates that the interaction between the adjacent OH groups is small. The OH frequencies of MXenes (see the supplemental data [77]) are comparable to the OH stretching modes of $H_2O$ (3585.5 cm$^{-1}$) or alcohol (3500-3700 cm$^{-1}$). This means that the OH bonds are fairly strong. These evidences imply that the hydrogen atoms may not be easily detached from the OH terminated MXenes. Moreover, recently experimentalists have successfully studied the OH stretching frequency modes (~3746 cm$^{-1}$) on silica at different temperatures up to 500 K [110].

The OH vibrational modes of MXenes are in the same range as that of silica [77]. Since the thermal stabilities of ceramics are typically higher than silica, it is expected the OH terminated MXenes to be also thermally stable at high temperatures around 500 K.

In order to compare the work function properties of the MXenes with other related materials, we have performed sets of calculations on the binary carbides (MC, M= Sc, Ti, Zr, Hf, V, Nb, Ta) and nitrides (M'N, M'= Ti, Zr, and Hf) as well as pure transition metals (Sc, Ti, Zr, Hf, V, Nb, Ta) functionalized with F, O, and OH. The results are summarized in the supplemental data [77]. We found only one experimental result on TiC that after oxygen exposure, its work function increases [111]. Such increase in the work function is also seen from our results. From our calculations, it is observed that in most of the MC and M'N systems, the F, O, or OH functionalization causes an increase in the work functions. Interestingly, the OH terminations on MC and M'N do not help to lower the work functions significantly. This is because the OH groups on MC and M'N are tilted. Hence, the polarity of the OH groups does not affect to lower the effective potentials and consequently the work functions of the MC and M'N systems. The work function properties of the functionalized transition metal surfaces are very similar to those for the functionalized MXenes [77]. In these systems, the OH groups stay perpendicular to the surface similar to the OH groups on MXene structures. Hence, the OH terminated transition metal surfaces obtain ultralow work functions similar to the OH terminated MXenes. However, owing to the ceramic nature of the MXenes, MXenes possess higher thermal stabilities than the pure transition metals and are superior for work function applications.

## 4. Summary:

By using a set of first-principles calculations, we studied large numbers of thin and thick MXenes: $M_2C$ and $M_{10}C_9$ (M= Sc, Ti, Zr, Hf, V, Nb, and Ta), as well as $M'_2N$ and $M'_{10}N_9$ (M'= Ti, Zr, and Hf) functionalized with F, OH, and O. Our calculations revealed that independently of the type of M elements, the OH terminated MXenes attain ultralow work functions. In addition, it was observed that the increase or decrease in the work functions of the MXenes upon F or O functionalization depends on the type of the M elements. Moreover, we systematically studied the thickness dependences of the work functions of the functionalized $Ti_{n+1}C_n$ (n=1-9) with F, OH and O, and found that regardless of the thickness, a family of $Ti_{n+1}C_n(OH)_2$ exhibits ultralow work function. We also found that the changes in the work functions of the functionalized MXenes are linearly correlated with the changes in the surface dipole moments. By considering the three important factors in tuning the surface dipole moments induced by i) the electron charge redistribution at the surface, ii) the surface structure relaxation, and iii) the polarity of attached chemical groups, we have successfully explained the different behavior of the work functions for the F, OH, and O terminated MXenes.

## Acknowledgment:

The authors thank the crew of the Center for Computational Materials Science at National Institute for Materials Science (NIMS) for the continuous support of the supercomputing facility.## References

[1] N. D. Lang and W. Kohn, *Theory of Metal Surfaces: Work Function,* Phys. Rev. B **3**, 1215 (1971).

[2] R. Smoluchowski, *Anisotropy of the Electronic Work Function of Metals*, Phys. Rev. **60**, 661 (1941).

[3] A. J. Bennett and C. B. Duke, *Influence of the Lattice Potential on the Electronic Structure of Metallic Interfaces: Dipole Effects*, Phys. Rev. **188**, 1060 (1969).

[4] J. Bardeen, *Theory of the Work Function. II. The Surface Double Layer*, Phys. Rev. **49**, 653 (1936).

[5] E. Wigner and J. Bardeen, *Theory of the Work Functions of Monovalent Metals*, Phys. Rev. **48**, 84 (1935).

[6] M. Methfessel, D. Hennig, and M. Scheffler, *Trends of the Surface Relaxations, Surface Energies, and Work Functions of the 4d Transition Metals*, Phys. Rev. B **46**, 4816 (1992).

[7] F. K. Schulte, *A Theory of Thin Metal Films: Electron Density, Potentials and Work Function*, Surf. Sci. 55, 427 (1976).

[8] P. J. Feibelman, *Static Quantum-Size Effects in Thin Crystalline, Simple-Metal Films*, Phys. Rev. B 27, 1991 (1983).12

[9] S. Ciraci and I. P. Batra, *Theory of the Quantum Size Effect in Simple Metals*, Phys. Rev. B 33, 4294 (1986).

[10] N. E. Singh-Miller and N. Marzari, *Surface Energies, Work functions, and Surface Relaxations of Low-Index Metallic Surfaces from First Principles*, Phys. Rev. B **80**, 235407 (2009).

[11] R. Ramprasad, P. von Allmen, and L. R. C. Fonseca, *Contributions to the work function: A Density-Functional Study of Adsorbates at Graphene Ribbon Edges*, Phys. Rev. B **60**, 6023 (1999).

[12] Y. Ando, Y. Gohda, and S. Tsuneyuki, *Dependence of the Schottky Barrier on the Work Function at Metal/SiON/SiC(0001) Interfaces Identified by First-Principles Calculations*, Surf. Sci. **606**, 1501 (2012).

[13] J. L. Freeouf and J. M. Woodall, *Schottky Barriers: An Effective work Function Model*, Appl. Phys. Lett. **39**, 727 (1981).

[14] L. W. Swanson and R. W. Strayer, *Field-Electron-Microscopy Studies of Cesium Layers on Various Refractory Metals: Work Function Change*, J. Chem. Phys. **48**, 2421 (1960).

[15] R. E. Weber and W. T. Peria, *Work Function and Structural Studies of Alkali-Covered Semiconductors*, Surf. Sci. **14**,13 (1969).

[16] J. Huang, Z. Xu, and Y. Yang, *Low-Work-Function Surface Formed by Solution-Processed and Thermally Deposited Nanoscale Layers of Cesium Carbonate*, Adv. Funct. Mater. **17**, 1966 (2007).

[17] N. D. Lang, *Theory of Work-Function Changes Induced by Alkali Adsorption*, Phys. Rev. B **4**, 4234 (1971).

[18] W. Ning, C. Kailai, and W. Dingsheng, *Work Function of Transition-Metal Surface with Submonolayer Alkali-Metal Coverage*, Phys. Rev. Lett. **56**, 2759 (1986).

[19] R. W. Gurney, *Theory of Electrical Double Layers in Adsorbed Films*, Phys. Rev. **47**, 479 (1935).

[20] M. Khazaei and Y. Kawazoe, *Effects of Cs Treatment on Field Emission Properties of Capped Carbon Nanotubes*, Surf. Sci. **601**, 1501 (2007).

[21] M. Khazaei, A. A. Farajian, H. Mizuseki, and Y. Kawazoe, *Cs Doping Effects on Electronic Structure of Thin Nanotubes*, Comput. Mater. Sci. **36**, 152 (2006).

[22] F. Gossenberger, T. Roman, K. Forster-Tonigold, and A. Groß, *Change of the Work Function of Platinum Electrodes Induced by Halide Adsorption*, Beilstein J. Nanotechnol. **5**, 152 (2014).

[23] W. E. Ford, D. Gao, N. Knorr, R. Wirtz, F. Scholz, Z. Karipidou, K. Ogasawara, S. Rosselli, V. Rodin, G. Nelles, and F. von Wrochem, *Organic Dipole Layers for Ultralow Work Function Electrodes*, ACS Nano 8, 9173 (2014).

[24] W. S. Williams, *Physics of Transition Metal Carbides*, Mater. Sci. Eng. A **105**, 1 (1988).

[25] M. Yoshitake, *Generic Trend of Work Functions in Transition-Metal Carbides and Nitrides*, J. Vac. Sci. Technol. A **32**, 061403 (2014).

[26] K. Kobayashi, *First-Principles Study of the Surface Electronic Structures of Transition Metal Carbides*, Jpn. J. Appl. Phys. **39**, 4311 (2000).

[27] K. Kobayashi, *First-Principles Study of the Electronic Properties of Transition Metal Nitride Surfaces*, Surf. Sci. **493**, 665 (2001).

[28] H. W. Hugosson, O. Eriksson, U. Jansson, A. V. Ruban, P. Souvatzis, and I. A. Abrikosov, *Surface Energies and Work Functions of the Transition Metal Carbides*, Surf. Sci. **557**, 243 (2004).

[29] M. Endo, H. Nakane, and H. Adachi, *Field Emission Characteristics of Transition-Metal Nitrides*, J. Vac. Sci. Technol. B **14**, 2114 (1996).

[30] M. Nagao, Y. Fujimori, Y. Gotoh, H. Tsuji, and J. Ishikawa, *Emission Characteristics of ZrN Thin Film Field Emitter Array Fabricated by Ion Beam Assisted Deposition Technique*, J. Vac. Sci. Technol. B **16**, 829 (1998).

[31] M. Naguib, M. Kurtoglu, V. Presser, J. Lu, J. Niu, M. Heon, L. Hultman, Y. Gogotsi, and M. W. Barsoum, *Two-Dimensional Nanocrystals Produced by Exfoliation of $Ti_3AlC_2$*, Adv. Mater. **23**, 4248 (2011).

[32] M. Naguib, O. Mashtalir, J. Carle, V. Presser, J. Lu, L. Hultman, Y. Gogotsi, and M. W. Barsoum, *Two-Dimensional Transition Metal Carbides*, ACS Nano **6**, 1322 (2012).

[33] M. Khazaei, M. Arai, T. Sasaki, M. Estili, and Y. Sakka, *Trends in Electronic Structures and Structural Properties of MAX Phases: A First-Principles Study on $M_2AlC$ (M = Sc, Ti, Cr, Zr, Nb, Mo, Hf, or Ta), $M_2AlN$, and Hypothetical $M_2AlB$ Phases*, J. Phys.: Condens. Matter **26**, 505503 (2014).

[34] M. W. Barsoum, *The $M_{N+1}AX_N$ Phases: A New Class of Solids: Thermodynamically Stable Nanolaminates*, Prog. Solid State Chem. **28**, 201 (2000).

[35] Z. M. Sun, *Progress in Research and Development on MAX Phases: A Family of Layered Ternary Compounds*, Int. Mater. Rev. **56**, 143 (2011).

[36] M. F. Cover, O. Warschkow, M. M. M. Bilek, and D. R. McKenzie, *A Comprehensive Survey of $M_2AX$ Phase Elastic Properties*, J. Phys.: Condens. Matter 21, 305403 (2009).

[37] M. Khazaei, M. Arai, T. Sasaki, M. Estili, and Y. Sakka, *The Effect of the Interlayer Element on the Exfoliation of Layered $Mo_2AC$ (A = Al, Si, P, Ga, Ge, As or In) MAX Phases into Two-Dimensional $Mo_2C$ Nanosheets*, Sci. Tech. Adv. Mater. 15, 014208 (2014).

[38] M. R. Lukatskaya, O. Mashtalir, C. E. Ren, Y. Dall'Agnese, P. Rozier, P. L. Taberna, M. Naguib, P. Simon, M. W. Barsoum, and Y. Gogotsi, *Cation Intercalation and High Volumetric Capacitance of Two-Dimensional Titanium Carbide*, Science **341**, 1502 (2013).

[39] M. Chidiu, M. R. Lukatskaya, M. -Q. Zhao, Y. Gogotsi, and M. W. Barsoum, *Conductive Two-Dimensional Titanium Carbide 'Clay' with High Volumetric Capacitance*, Nature **516**, 78 (2014).

[40] M. Naguib, J. Halim, J. Lu, K. M. Cook, L. Hultman, Y. Gogotsi, and M. W. Barsoum, *New Two-Dimensional Niobium and Vanadium Carbides as Promising Materials for Li-Ion Batteries*, J. Am. Chem. Soc. **135**, 15966 (2013).

[41] Y. Xie, Y. Dall'Agnese, M. Naguib, Y. Gogotsi, M. W. Barsoum, H. L. Zhuang, and P. R. C. Kent, *Prediction and Characterization of MXene Nanosheet Anodes for Non-Lithium-Ion Batteries*, ACS Nano **8**, 9606 (2014).

[42] Q. Tang, Z. Zhou, and P. Shen, *Are MXenes Promising Anode Materials for Li Ion Batteries? Computational Studies on Electronic Properties and Li Storage Capability of $Ti_3C_2$ and $Ti_3C_2X_2$ (X = F, OH) Monolayer*, J. Am. Chem. Soc. **134**, 16909 (2012).

[43] C. Eames and M. S. Islam, *Ion Intercalation into Two-Dimensional Transition-Metal Carbides: Global Screening for New High-Capacity Battery Materials*, J. Am. Chem. Soc. **136**, 16270 (2014).

[44] Dequan Er, Junwen Li, Michael Naguib, Yury Gogotsi, and Vivek B. Shenoy, *$Ti_3C_2$ MXene as a High Capacity Electrode Material for Metal (Li, Na, K, Ca) Ion Batteries*, ACS Appl. Mater. Interfaces **6**, 11173 (2014).

[45] J. Hu, B. Xu, C. Ouyang, S. A. Yang, and Y. Yao, *Investigations on $V_2C$ and $V_2CX_2$ (X = F, OH) Monolayer as a Promising Anode Material for Li Ion Batteries from First-Principles Calculations*, J. Phys. Chem. C **118**, 24274 (2014).

[46] E. Yang, H. Ji, J. Kim, H. Kim, and Y. Jung, *Exploring the Possibilities of Two-Dimensional Transition Metal Carbides as Anode Materials for Sodium Batteries*, Phys. Chem. Chem. Phys. **17**, 5000 (2015).

[47] J. Halim, M. Lukatskaya, K. M. Cook, J. Lu, C. R. Smith, L. Å. Näslund, S. J. May, L. Hultman, Y. Gogotsi, P. Eklund, and M. W. Barsoum, *Transparent Conductive Two-Dimensional Titanium Carbide Epitaxial Thin Films*, Chem. Mater. **26**, 2374 (2014).

[48] X. Zhang, J. Xu, H. Wang, J. Zhang, H. Yan, B. Pan, J. Zhou, and Y. Xie, *Ultrathin Nanosheets of MAX Phases with Enhanced Thermal and Mechanical Properties in Polymeric Compositions: $Ti_3Si_{0.75}Al_{0.25}C_2$*, Angew. Chem. Int. Ed. **52**, 4361 (2013).

[49] Q. Peng, J. Guo, Q. Zhang, J. Xiang, B. Liu, A. Zhou, R. Liu, and Y. Tian, *Unique Lead Adsorption Behavior of Activated Hydroxyl Group in Two-Dimensional Titanium Carbide*, J. Am. Chem. Soc. **136**, 4113 (2014).

[50] J. Chen, K. Chen, D. Tong, Y. Huang, J. Zhang, J. Xue, Q. Huang and T. Chen, *$CO_2$ and Temperature Dual Responsive "Smart" MXene Phases*, Chem. Commun. **51**, 314 (2015).

[51] O. Mashtalir, K. M. Cook, V. N. Mochalin, M. Crowe, M. W. Barsoum, and Y. Gogotsi, *Dye Adsorption and Decomposition on Two-Dimensional Titanium Carbide in Aqueous Media*, J. Mater. Chem. A **2**, 14334 (2014).

[52] M. Khazaei, M. Arai, T. Sasaki, C. -Y. Chung, N. S. Venkataramanan, M. Estili, Y. Sakka, and Y. Kawazoe, *Novel Electronic and Magnetic Properties of Two-Dimensional Transition Metal Carbides and Nitrides*, Adv. Funct. Mater. 23, 2185 (2013).

[53] M. Kurtoglu, M. Naguib, Y. Gogotsi, and M. W. Barsoum, *First Principles Study of Two-Dimensional Early Transition Metal Carbides*, MRS Commun. 2, 133 (2012).

[54] Y. Xie and P. R. C. Kent, *Hybrid Density Functional Study of Structural and Electronic Properties of Functionalized $Ti_{n+1}X_n$ (X=C, N) Monolayers*, Phys. Rev. B 87, 235441 (2013).

[55] A. N. Enyashin and A. L. Ivanovskii, *Structural and Electronic Properties and Stability of MXenes $Ti_2C$ and $Ti_3C_2$ Functionalized by Methoxy Groups*, J. Phys. Chem. C **117**, 13637 (2013).

[56] S. Zhao, W. Kang, and J. Xue, *MXene Nanoribbons*, J. Mater. Chem. C **3**, 879 (2015).

[57] Z. Ma, Z. Hu, X. Zhao, Q. Tang, D. Wu, Z. Zhou, and L. Zhang, *Tunable Band Structures of Heterostructured Bilayers with Transition-Metal Dichalcogenide and MXene Monolayer*, J. Phys. Chem. C **118**, 5593 (2014).

[58] Y. Lee, S. B. Cho, and Y. C. Chung, *Tunable Indirect to Direct Band Gap Transition of Monolayer $Sc_2CO_2$ by the Strain Effect*, ACS Appl. Mater. Interfaces **6**, 14724 (2014).

[59] Y. Lee, S. B. Cho, and Y. -C. Chung, *Achieving a Direct Band Gap in Oxygen Functionalized-Monolayer Scandium Carbide by Applying an Electric Field*, Phys. Chem. Chem. Phys. **16**, 26273 (2014).

[60] X. Li, Y. Dai, Y. Ma, Q. Liu, and B. Huang, *Intriguing Electronic Properties of Two-Dimensional $MoS_2/TM_2CO_2$ (TM = Ti, Zr, or Hf) Hetero-Bilayers: Type-II Semiconductors with Tunable Band Gaps*, Nanotechnology 26, 135703 (2015).

[61] X. -F. Yu, J. -B. Cheng, Z. -B. Liu, Q. -Z. Li, W. -Z. Li, X. Yang, and B. Xia, *The Band Gap Modulation of Monolayer $Ti_2CO_2$ by Strain*, RSC Adv. 5, 30438 (2015).

[62] L. -Y. Gan, D. Huang, and U. Schwingenschlögl, *Oxygen adsorption and dissociation during the oxidation of monolayer $Ti_2C$*, J. Mater. Chem. A **1**, 13672 (2013).

[63] H. Weng, A. Ranjbar, Y. Liang, Z. Song, M. Khazaei, S. Yunoki, M. Arai, Y. Kawazoe, Z. Fang, and X. Dai, *Large-Gap Two-dimensional Topological Insulator in Oxygen Functionalized MXene*, arXiv:1507.01172 [cond-mat.mtrl-sci].

[64] H. Fashandi, V. Ivády, P. Eklund, A. Lloyd Spetz, M. I. Katsnelson, and I. A. Abrikosov, *Dirac Points with Giant Spin-Orbit Splitting in the Electronic Structure of Two-Dimensional Transition Metal Carbides*, arXiv:1506.05398 [cond-mat.mtrl-sci].

[65] S. Zhao, W. Kang, and J. Xue, *Manipulation of Electronic and Magnetic Properties of $M_2C$ (M = Hf, Nb, Sc, Ta, Ti, V, Zr) Monolayer by Applying Mechanical Strains*, Appl. Phys. Lett. **104**, 133106 (2014).

[66] N. J. Lane, M. W. Barsoum, and J. M. Rondinelli, *Correlation Effects and Spin-Orbit Interactions in Two-Dimensional Hexagonal 5d Transition Metal Carbides, $Ta_{n+1}C_n$ (n = 1,2,3)*, Europhys. Lett. **15**, 014208 (2014).

[67] H. Lashgari, M. R. Abolhassani, A. Boochani, S. M. Elahi, and J. Khodadadi, *Electronic and Optical Properties of 2D Graphene-Like Compounds Titanium Carbides and Nitrides: DFT Calculations*, Solid State Commun. 195, 61 (2014).

[68] M. Khazaei, M. Arai, T. Sasaki, M. Estili, and Y. Sakka, *Two-Dimensional Molybdenum Carbides: Potential Thermoelectric Materials of the MXene Family*, Phys. Chem. Chem. Phys. 16, 7841 (2014).

[69] X. -F. Yu, Y. Li, J. -B. Cheng, Z. -B. Liu, Q. -Z. Li, W. -Z. Li, X. Yang, and B. Xiao, *Monolayer $Ti_2CO_2$: A Promising Candidate for $NH_3$ Sensor or Capturer with High Sensitivity and Selectivity*, ACS Appl. Mater. Interfaces **7**, 13707 (2015).

[70] L. -Y. Gan, Y. -J. Zhao, D. Huang, and U. Schwingenschlögl, *First-Principles Analysis of $MoS_2/Ti_2C$ and $MoS_2/Ti_2CY_2$ (Y=F and OH) all-2D Semiconductor/Metal Contacts*, Phys. Rev. B **87**, 245307 (2013).

[71] H. Zhao, C. Zhang, S. Li, W. Ji, and P. Wang, *First-Principles Design of Silicene/$Sc_2CF_2$ Heterojunction as a Promising Candidate for Field Effect Transistor*, J. Appl. Phys. **117**, 085306 (2015).

[72] Y. Lee, Y. Hwang, and Y. C. Chung, *Achieving Type I, II, and III Heterojunctions Using Functionalized MXene*, ACS Appl. Mater. Interfaces **7**, 7163 (2015).

[73] J. P. Perdew, K. Burke, and M. Ernzerhof, *Generalized Gradient Approximation Made Simple*, Phys. Rev. Lett. 77, 3865 (1996).

[74] M. Methfessel and A. T. Paxton, *High-Precision Sampling for Brillouin-Zone Integration in Metals*, Phys. Rev. B 40 3616 (1989).

[75] H. J. Monkhorst and J. D. Pack, *Special Points for Brillouin-Zone Integrations*, Phys. Rev. B 13, 5188 (1976).

[76] G. Kresse and J. Furthmüller, *Efficiency of Ab-Initio Total Energy Calculations for Metals and Semiconductors Using a Plane-Wave Basis Set*, Comput. Mater. Sci. **6**, 15 (1996).

[77] See Supplemental Material at [*URL will be inserted by publisher*] for total energies of the optimized structures; electronic structures of $Ti_{n+1}C_n$, $Ti_{n+1}C_nF_2$, $Ti_{n+1}C_n(OH)_2$, $Ti_{n+1}C_nO_2$ (n = 1-9), $M_{10}C_9$ (M= Sc, Ti, Zr, Hf, V, Nb, Ta, and Mo), $M'_2N$ and $M'_{10}N_9$ (M'= Ti, Zr, and Hf); the effect of unit cell size on the work functions; the interatomic elastic constant ($C_{33}$) of the $Sc_{10}C_9$ and $Hf_{10}C_9$; and phonon dispersions of $Sc_2C(OH)_2$, $Ti_2C(OH)_2$, $Zr_2C(OH)_2$, $Zr_2N(OH)_2$, $Hf_2C(OH)_2$, $Hf_2N(OH)_2$, and $Ta_2C(OH)_2$; some benchmarks for work function calculations; work function properties of pure transition metal surfaces as well as binary carbides (MC) and nitrides (MN) functionalized with F, OH, and O; calculated detachment energies of the hydrogen atoms from the OH terminated MXenes.

[78] A. Togo, F. Oba, and I. Tanaka, *First-Principles Calculations of the Ferroelastic Transition between Rutile-Type and $CaCl_2$-type SiO2 at High Pressures*, Phys. Rev. B **78**, 134106 (2008).

[79] D. L. Price, B. R. Cooper, and J. M. Wills, *Full-Potential Linear-Muffin-Tin-Orbital Study of Brittle Fracture in Titanium Carbide*, Phys. Rev. B **46**, 11368 (1992).

[80] E. Wimmer, A. Neckel, and A. J. Freeman, *TiC(001) Surface: All-Electron Local-Density-Functional Study*, Phys. Rev. B **31**, 2370 (1985).

[81] David L. Price, Bernard R. Cooper, and John M. Wills, *Effect of Carbon Vacancies on Carbide Work Functions*, Phys. Rev. B **48**, 15311 (1993).

[82] H. L. Skriver and N. M. Rosengaard, *Surface Energy and Work Function of Elemental Metals*, Phys. Rev. B **46**, 7157 (1992).

[83] H. B. Michaelson, *The Work Function of the Elements and Its Periodicity*, J. Appl. Phys. **48**, 4729 (1977).

[84] F. Viñes, C. Sousa, P. Liu, J. A. Rodriguez, and F. Illas, *A Systematic Density Functional Theory Study of the Electronic Structure of Bulk and (001) Surface of Transition-Metals Carbides*, J. Chem. Phys. **122**, 174709 (2005).

[85] H. W. Hugosson, O. Eriksson, U. Jansson, A.V. Ruban, P. Souvatzis, and I.A. Abrikosov, *Surface Energies and Work Functions of the Transition Metal Carbides*, Surf. Sci. **557**, 243 (2004).

[86] K. Kobayashi, *First-Principles Study of the Electronic Properties of Transition Metal Nitride Surfaces*, Surf. Sci. **493**, 665 (2001).

[87] C. Oshima, M. Aono, S. Zaima, Y. Shibata, and S. Kawai, The Surface Properties of TiC(001) and TiC(111) Surfaces, J. Less-Common Metal **82**, 69 (1981).

[88] Y. Saito, S. Kawata, H. Nakane, and H. Adachi, *Emission Characteristics of Niobium Nitride Field Emitters*, Appl. Surf. Sci. **146**, 177 (1999).

[89] T. Aizawa, Rep. of National Institute for Research in Organic Materials, vol. **81**, p. 27, 1994.

[90] P. A. P. Lindberg and L. I. Johansson, *Work Function and Reactivity of Some Crystal Faces of Substoichiometric Transition-Metal Carbides*, Surf. Sci. **194**, 199 (1988).

[91] G. R. Gruzalski, S. -C. Lui, and D.M. Zehner, *Work-Function Changes Accompanying Changes in Composition of (100) Surfaces of $HfC_x$ and $TaC_x$*, Surf. Sci. Lett. **239**, L517 (1990).

[92] T. C. Leung, C. L. Kao, W. S. Su, Y. J. Feng, and C. T. Chan, *Relationship between Surface Dipole, Work Function and Charge Transfer: Some Exceptions to an Established Rule*, Phys. Rev. B **68**, 195408 (2003).

[93] J. P. Palmquist, S. Li, P. O. Å. Persson, J. Emmerlich, O. Wilhelmsson, H. Högberg, M. I. Katsnelson, B. Johansson, R. Ahuja, O. Eriksson, L. Hultman, and U. Jansson, *$M_{n+1}AX_n$ Phases in the Ti-Si-C system Studied by Thin-Film Synthesis and Ab Initio Calculations*, Phys. Rev. B **70**, 165401 (2004).

[94] M. W. Barsoum, H. -I. Yoo, I. K. Polushina, V. Yu. Rud', Yu. V. Rud', and T. El-Raghy, *Electrical Conductivity, Thermopower, and Hall Effect of $Ti_3AlC_2$, $Ti_4AlN_3$, and $Ti_3SiC_2$*, Phys. Rev. B **62**, 10194 (2000).

[95] H. Högberg, P. Eklund, J. Emmerlich, J. Birch, L. Hultman, *Epitaxial $Ti_2GeC$, $Ti_3GeC_2$, and $Ti_4GeC_3$ MAX-Phase Thin Films Grown by Magnetron Sputtering*, J. Mater. Res. **20**, 779 (2005).

[96] C. Hu, J. Zhang, J. Wang, F. Li, J. Wang, and Y. Zhou, *Crystal Structure of $V_4AlC_3$: A New Layered Ternary Carbide*, J. Am. Ceram. Soc. **91**, 636 (2008).

[97] C. Hu, F. Li, L. He, M. Liu, J. Zhang, J. Wang, Y. Bao, J. Wang, and Y. Zhou, *In Situ Reaction Synthesis, Electrical and Thermal, and Mechanical Properties of $Nb_4AlC_3$*, J. Am. Ceram. Soc. **91**, 2258 (2008).

[98] B. Manoun, S. K. Saxena, T. El-Raghy, and M. W. Barsoum, *High-Pressure X-Ray Diffraction Study of $Ta_4AlC_3$*, Appl. Phys. Lett. **88**, 201902 (2006).

[99] J. Etzkorn, M. Ade, D. Kotzott, M. Kleczek, and H. Hillebrecht, *$Ti_2GaC$, $Ti_4GaC_3$ and $Cr_2GaC$-Synthesis, Crystal Growth and Structure Analysis of Ga-Containing MAX-Phases $M_{n+1}GaC_n$ with M= Ti, Cr and n=1, 3*, J. Solid State Chem. **182**, 995 (2009).

[100] L.Y. Zheng, J. M. Wang, X. P. Lu, F. Z. Li, J. Y. Wang, and Y. C. Zhou, *$(Ti_{0.5}Nb_{0.5})_5AlC_4$: A New-Layered Compound Belonging to MAX Phase*, J. Am. Ceram. Soc. **93**, 3068 (2010).

[101] Z. J. Lin, M. J. Zhuo, Y. C. Zhou, M. S. Li, and J. Y. Wang, *Microstructures and Theoretical Bulk Modulus of Layered Ternary Tantalum Aluminum Carbides*, J. Am. Ceram. Soc. **89**, 3765 (2006).

[102] J. Zhang, B. Liu, J. Y. Wang, and Y. C. Zhou, *Low-Temperature Instability of $Ti_2SnC$: A Combined Transmission Electron Microscopy, Differential Scanning Calorimetry, and X-Ray Diffraction Investigations*, J. Mater. Res. **24**, 39 (2009).

[103] I. P. Batra, S. Ciraci, G. P. Srivastava, J. S. Nelson, and C. Y. Fong, *Dimensionality and Size Effects in Simple Metals*, Phys. Rev. B **34**, 8246 (1986).

[104] N. Jiao, C. He, P. Zhou, C. X. Zhang, H. P. Xiao, and L. Z. Sun, *Surface Work Function of Chemically Derived Graphene: A first-Principles Study*, Phys. Lett. A **377**, 1760 (2013).

[105] N. Jiao, C. He, P. Zhou, C. X. Zhang, X. Peng, K. W. Zhang, and L. Z. Sun, *Modulation Effect of Hydrogen and Fluorine Decoration on the Surface Work Function of BN Sheets*, AIP Adv. **2**, 022125 (2012).

[106] K. T. Chan, J. B. Neaton, and M. L. Cohen, *First-Principles Study of Metal Adatom Adsorption on Graphene*, Phys. Rev. B **77**, 235430 (2008).

[107] P. C. Rusu and G. Brocks, *Surface Dipoles and Work Functions of Alkylthiolates and Fluorinated Alkylthiolates on Au(111)*, J. Phys. Chem. B, **110**, 22628 (2006).

[108] V. Mauchamp, M. Bugnet, E. P. Bellido, G. A. Botton, P. Moreau, D. Magne, M. Naguib, T. Cabioc'h, and M. W. Barsoum, *Enhanced and Tunable Surface Plasmons in Two-Dimensional $Ti_3C_2$ Stacks: Electronic Structure Versus Boundary Effects*, Phys. Rev. B **89**, 235428 (2014).

[109] T. Hu, J. Wang, H. Zhang, Z. Li, M. Hu, and X. Wang, *Vibrational Properties of $Ti_3C_2$ and $Ti_3C_2T_2$ (T = O, F, OH) Nonosheets by First-Principles Calculations: A Comparative Study*, Phys. Chem. Chem. Phys. **17**, 9997 (2015).

[110] H. Yamazaki, H. Shima, E. Yoda, and J. N. Konda, *Estimation of the Real Temperature of Samples in IR Cell Using OH Frequency of Silica*, Surf. Interface Anal. **47**, 166 (2015).

[111] P. A. P. Lindberg and L. I. Johansson, *Work Function and Reactivity of Some Crystal Faces of Substoichiometric Transition-Metal Carbides*, Surf. Sci. **194**, 199 (1988).

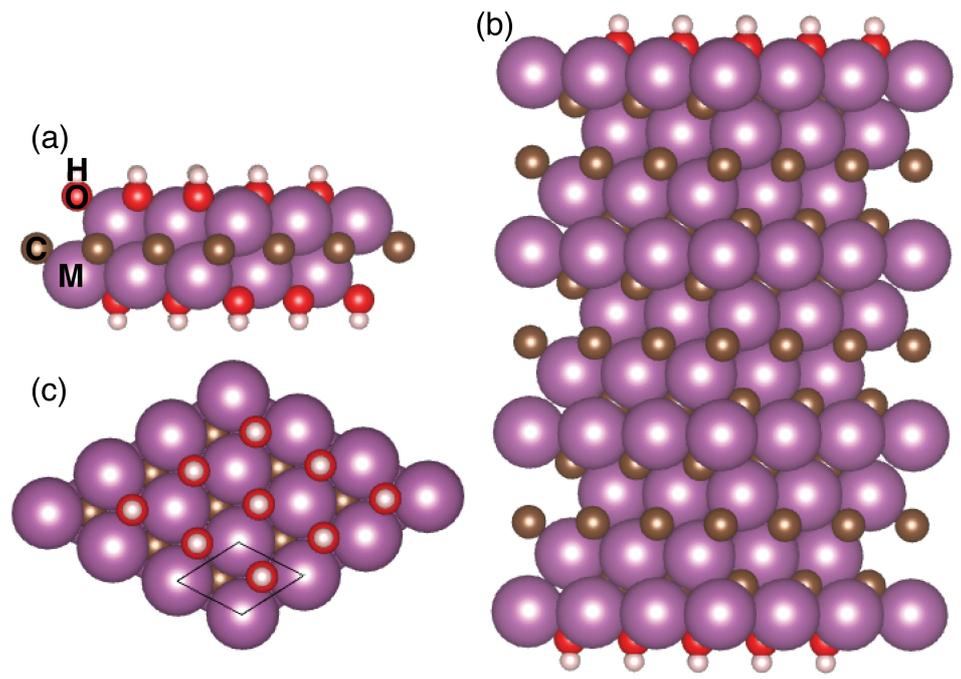

**FIG. 1**. The side views of 3×3 models of (a) the thinnest ($M_2C$) and (b) the thickest ($M_{10}C_9$) MXenes functionalized with OH chemical groups, which were considered in this study. (c) The top views of (a) and (b).

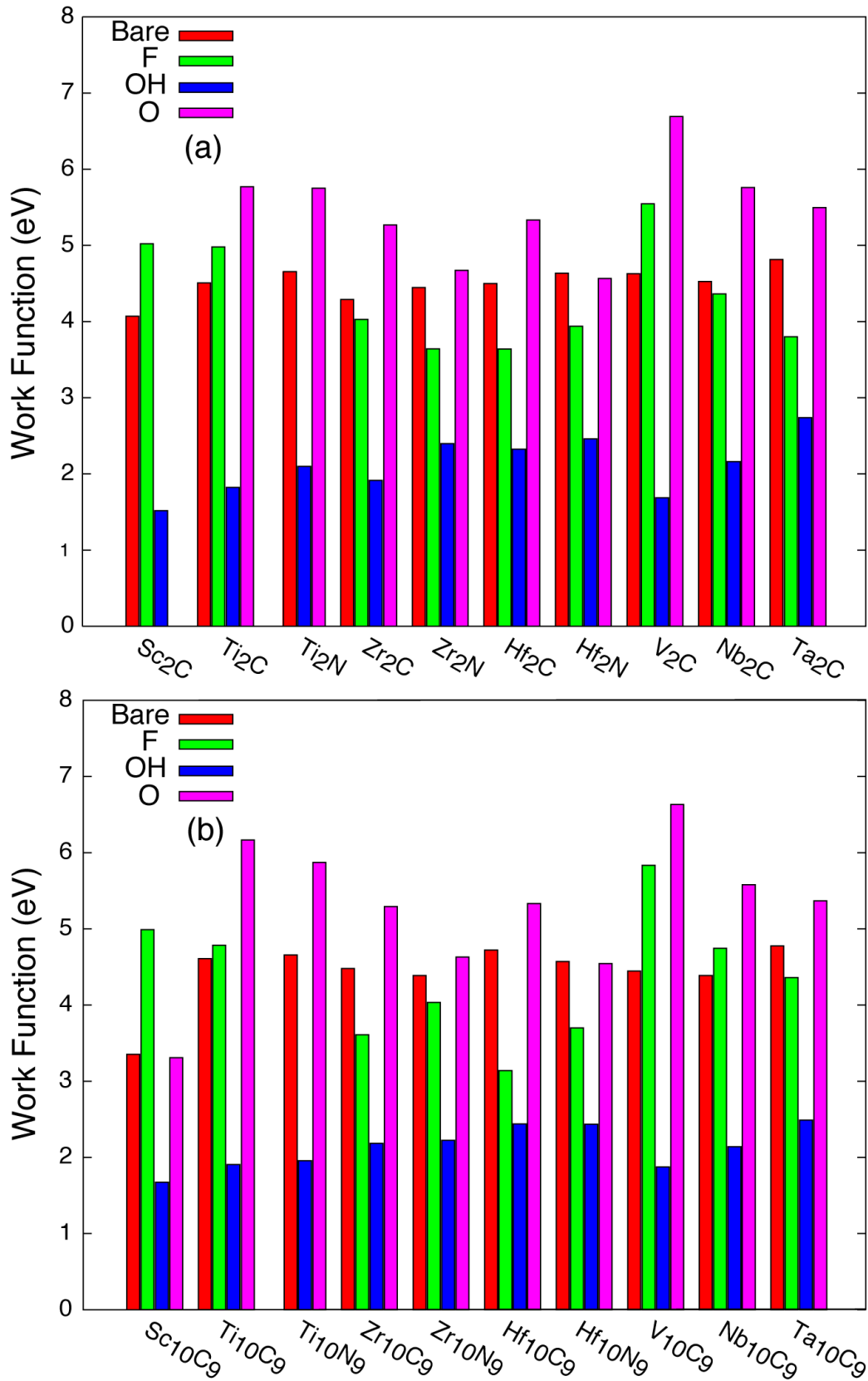

FIG. 2. The work functions of (a) M$_2$C and M'$_2$N, and (b) M$_{10}$C$_9$ and M'$_{10}$N$_9$ (M= Sc, Ti, Zr, Hf, V, Nb, Ta ; M'= Ti, Zr, Hf) functionalized with F, OH, and O.

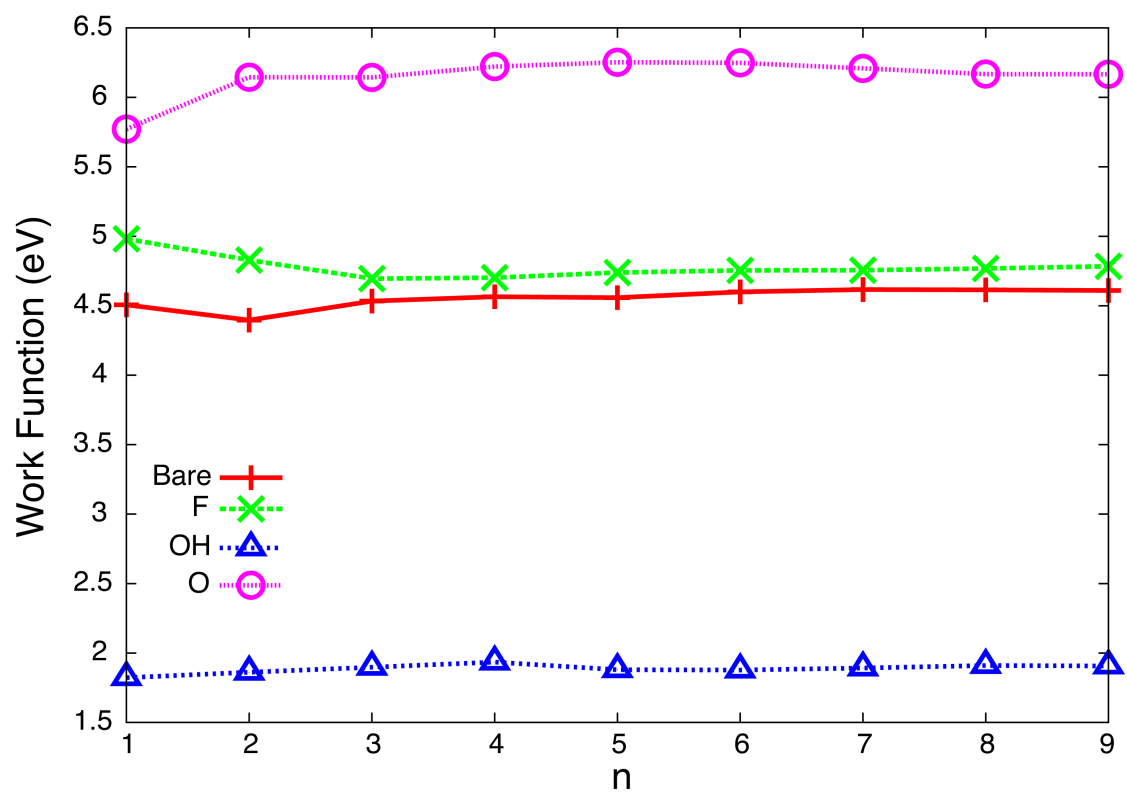

**FIG. 3**. Work functions of bare and functionalized $Ti_{n+1}C_n$ with F, OH, and O chemical groups.

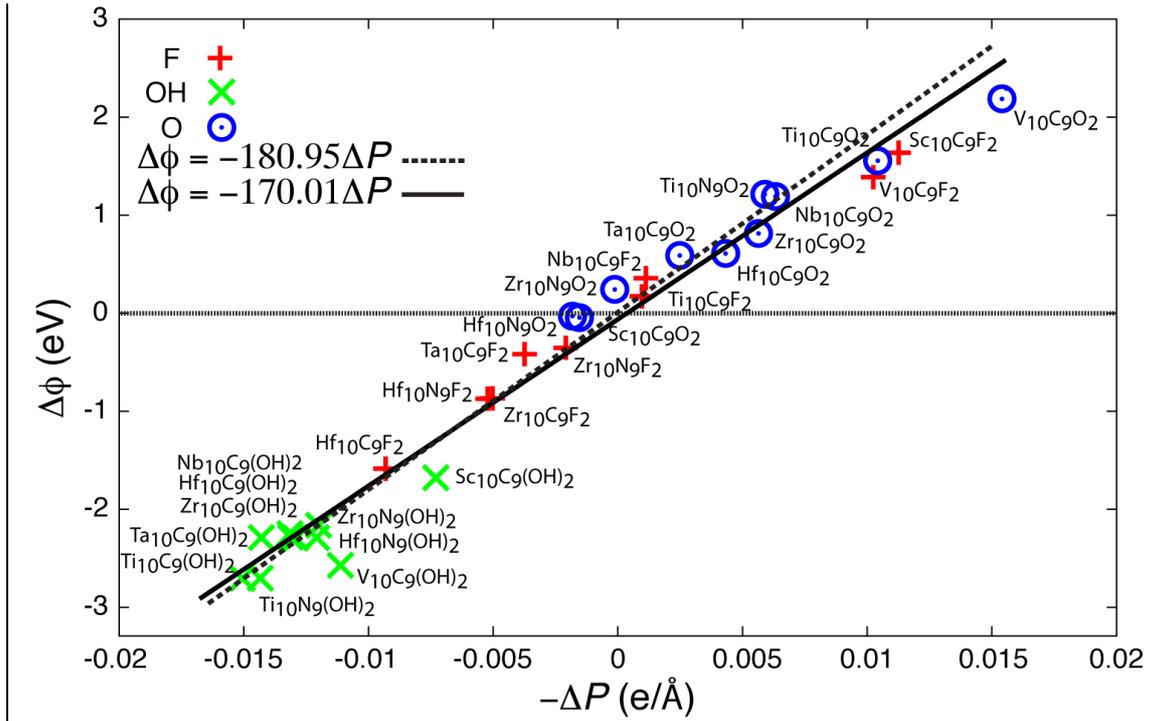

**FIG. 4**. The changes in the work functions (ΔΦ) of functionalized $M_{10}C_9$ and $M'_{10}N_9$ (M= Sc, Ti, Zr, Hf, V, Nb, Ta ; M'= Ti, Zr, Hf) with F, OH, and O as a function of the changes in the surface dipole moments (ΔP). The solid line (ΔΦ = -170.01ΔP) indicates the linear fit to the results. The dashed line (ΔΦ = -180.95ΔP) indicates the linear relation reported previously for bulk three-dimensional systems [92].

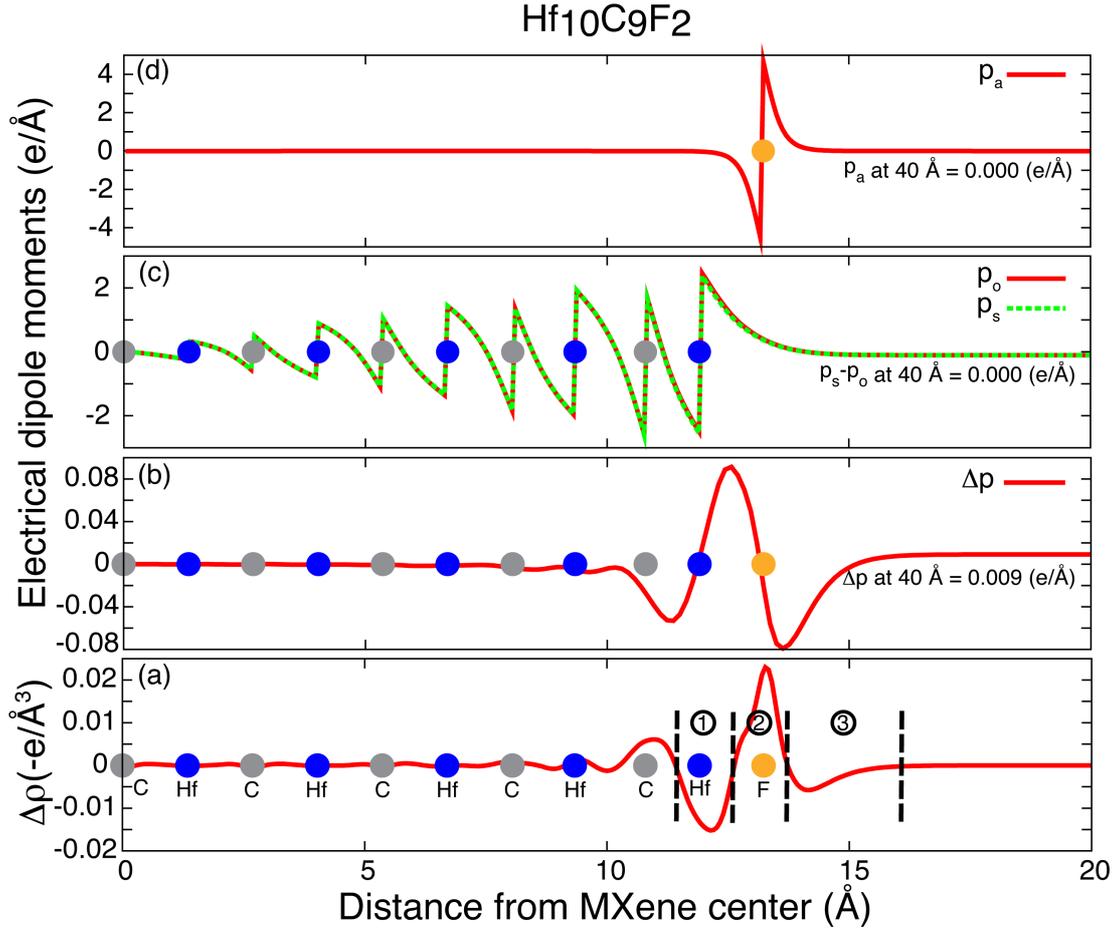

**FIG. 5**. (a) The transferred electron charge density ($\Delta\rho$) between adsorbate (F) and substrate ($Hf_{10}C_9$). The vertical dashed lines indicate the encompassing regions of positive (① and ③) and negative charges (②) nearby the surfaces. (b)-(d) the contributions to the total surface dipole moments ($\Delta P$): (b) the dipole moment due to the transferred electron charges between surface and the adsorbate ($\Delta p$), (c) the dipole moment of the optimized bare surface ($p_o$) and the dipole moment of the bare surface without optimization ($p_s$) (see the text), and (d) the dipole moment of adsorbate ($p_a$). $\Delta\rho$, $\Delta p$, $p_o$, $p_s$, and $p_a$ have been calculated as a function of z (along the normal surface) from the center of MXenes to the end of the cell (40 Å), but the results are shown for distances between 0 and 20 Å only. The respective quantities at 40 Å are also indicated in (b)-(d). The circles indicate the positions of the layers and their composed elements.

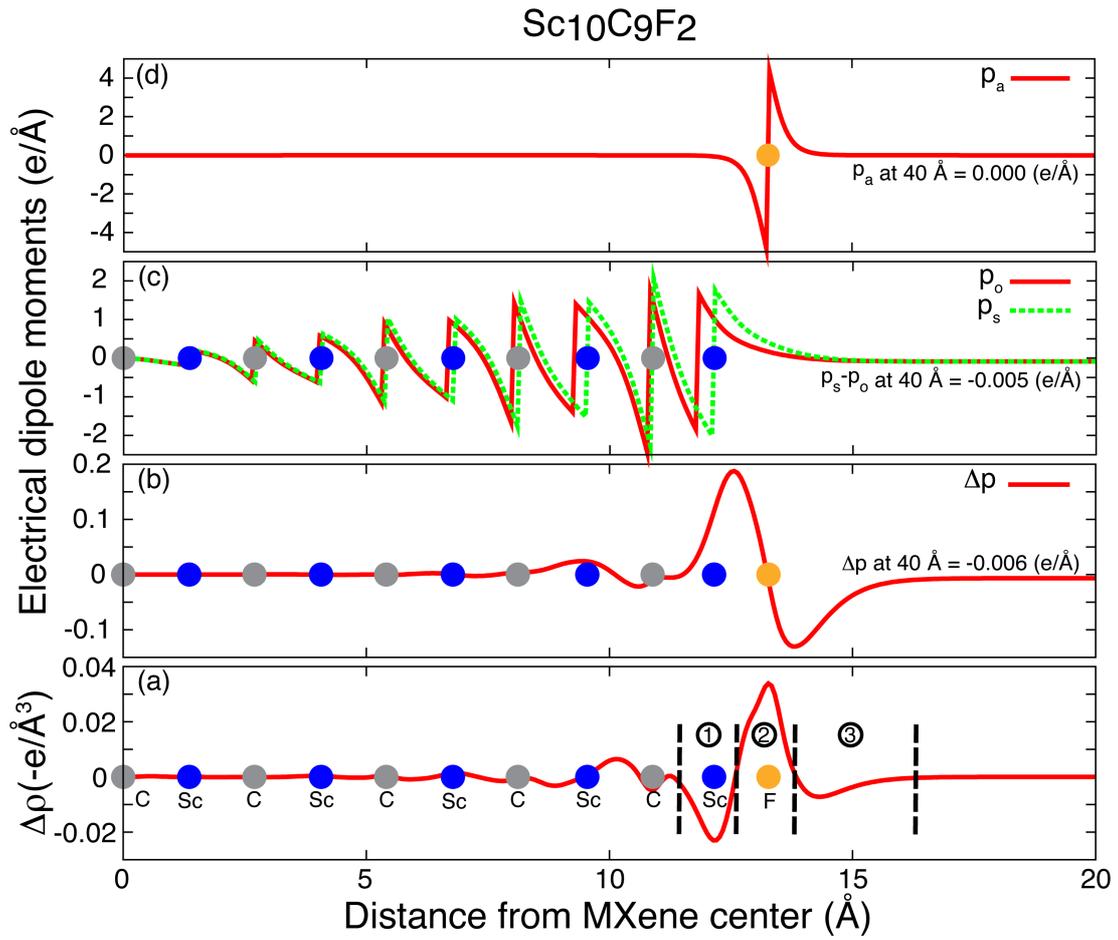

**FIG. 6**. The same as Fig. 5 but for $Sc_{10}C_9F_2$.

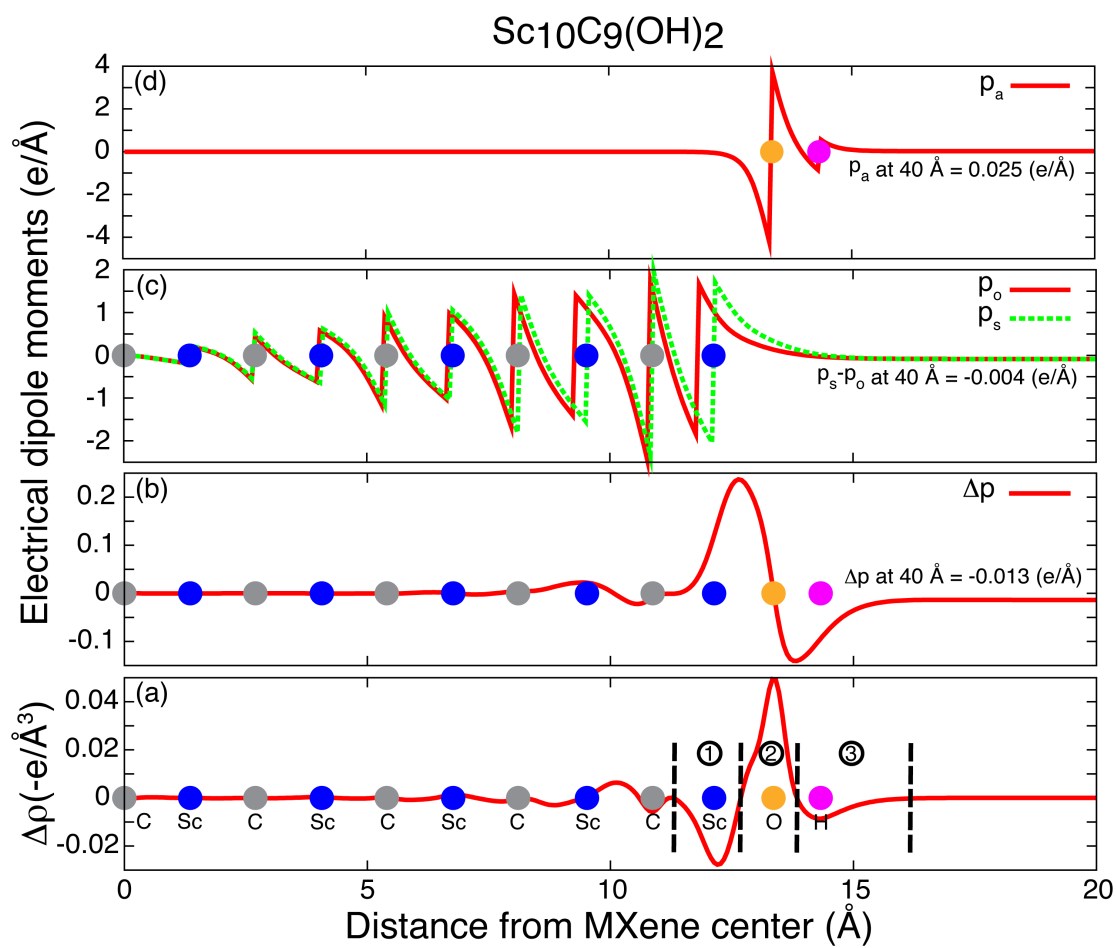

FIG. 7. The same as Fig. 5 but for $Sc_{10}C_9(OH)_2$.

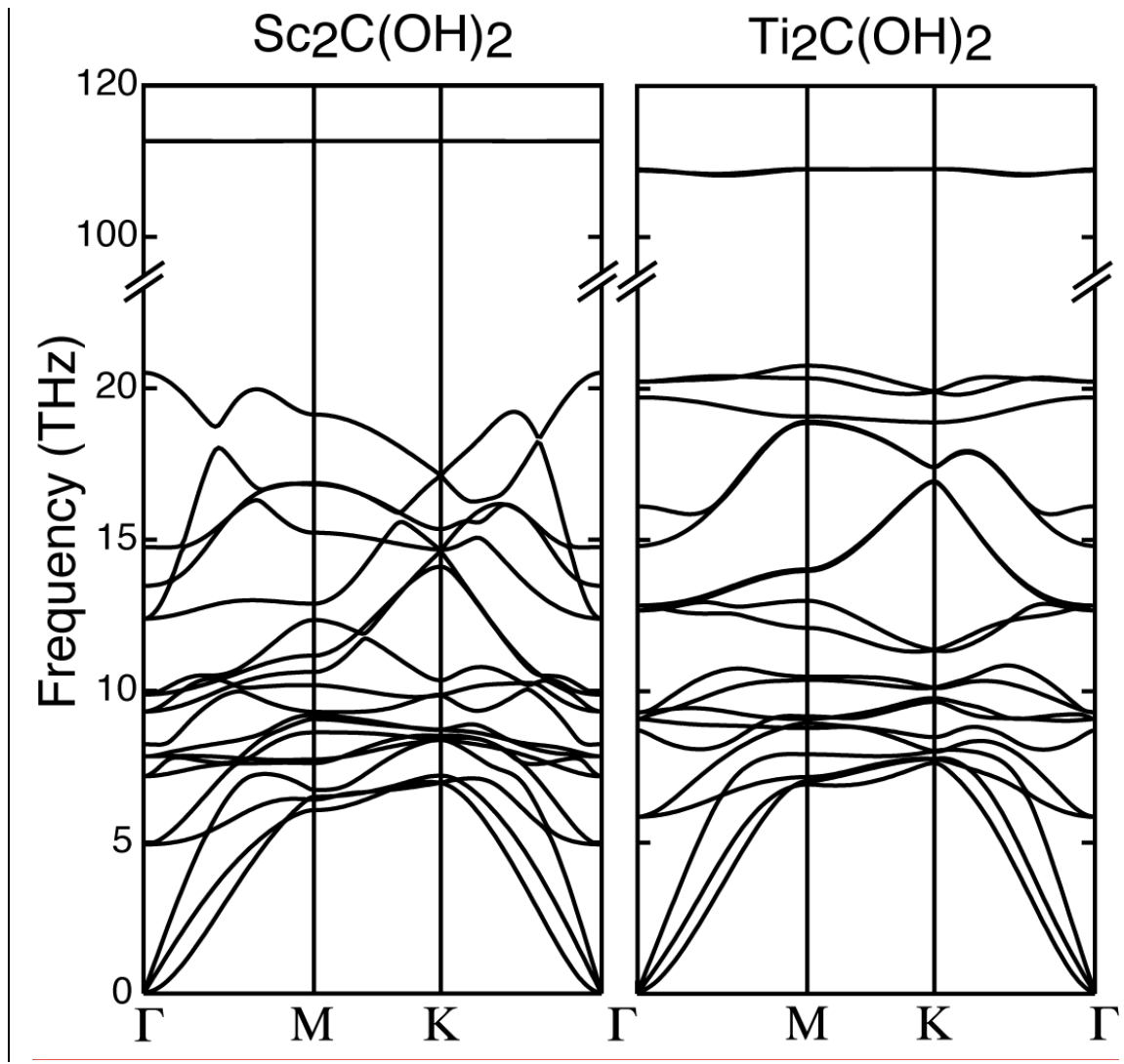

**FIG. 8**. Phonon dispersions of $Sc_2C(OH)_2$ and $Ti_2C(OH)_2$.